\documentclass[12pt]{article}

\usepackage{amsmath}
\usepackage{amsfonts}
\usepackage{bm}
\usepackage{multirow}
\usepackage{natbib}
\usepackage{pdfpages}
\usepackage[margin=.75in]{geometry}

\title{One-inflated zero-truncated Poisson and negative binomial regression models}
\author{Ryan T. Godwin\footnote{\noindent email: ryan.godwin@umanitoba.ca; address: University of Manitoba, 646 Fletcher Argue Building, Winnipeg, MB, R3T5V5, Canada}}
\date{}

\begin{document}

\maketitle

\abstract{
The workhorse model for zero-truncated count data ($y=1,2,\dots$) is the zero-truncated negative binomial (ZTNB) model. We find it should seldom be used. Instead, we recommend the one-inflated zero-truncated negative binomial (OIZTNB) model developed here. Zero-truncated count data often contain an excess of 1s, leading to bias and inconsistency in the ZTNB model. The importance of the OIZTNB model is apparent given the obvious presence of one-inflation in four datasets that have traditionally championed the standard ZTNB. We provide estimation, marginal effects, and a suite of accompanying tools in the \texttt{R} package \texttt{oneinfl}, available on CRAN.
}

\section{Introduction} \label{sec:intro}

The zero-truncated negative binomial (ZTNB) model is the workhorse model for when the outcome variable is a non-zero count ($y=1,2,...$). Yet, we find that ZTNB is misspecified in nearly every instance, due to one-inflation, which leads to biased and inconsistent estimators for the parameters of the model. It is common to find a preponderance of 1 counts in data, but distributions that allow for one-inflation have only recently been developed. While gaining popularity in biometrics, criminology and epidemiology, one-inflation has not yet extended into economics and other social sciences, perhaps due to a lack of an available regression model\footnote{A brief attempt was made in an unpublished conference paper by \citep{hassanzadeh2014one}}. To rectify this we develop the one-inflated zero-truncated negative binomial (OIZTNB) and one-inflated positive Poisson (OIPP) regression models. To illustrate their importance, we apply them to four common datasets traditionally used to highlight and exemplify the standard ZTNB, finding the presence of one-inflation in each case. We will argue that one-inflation should be the norm when data is collected administratively, and therefore recommend the estimation of OIZTNB instead of ZTNB, for most truncated data.

Truncated count models are used in several areas of economics. In health economics, individual health care usage is often measured through doctor visits or days in a hospital. Patients do not enter the administrative sample until they make their first visit, so 0 counts are not observed. In tourism economics and recreation demand, the number of times an individual visits a site is often collected on-site making 0 counts impossible. In labour economics, the number of times a worker becomes unemployed or changes jobs is often zero-truncated when the data are collected administratively. Further examples of interest to social scientists include the number of arrests or incarcerations, the number of patents filed by a firm, or the number of insurance claims.

Truncation occurs if the observational apparatus only becomes active with the occurrence of one event \citep{johnson1969distributions}. One-inflation occurs if the observational apparatus has an effect on that which is being observed, or on some associated party. It is usually not only the researcher that gains information at the first event. Revelations made at first visit disrupt the probability of a 1 count, but subsequent visits do not garner additional information, and the distortion to the counting distribution is confined to the 1s. A doctor may learn that a patient is healthy and need not return after the first visit. A consumer may learn that they hate the experience and only do it once. A worker may have lower utility than they expected due to unemployment, and learn to avoid subsequent unemployment spells. In situations where information is learned at first visit, but not from subsequent visits, one-inflation (or deflation) is likely to occur.

The recent attention to one-inflated distributions in biology, epidemiology, and criminology, is due to population size estimation. A truncated count model is fit to ``capture-recapture'' data, in order to infer the number of missing 0s, thus obtaining an estimate for the entire population. One-inflation has been found for many such data. For example, consider the number of times a prostitute is arrested. \cite{rossmo1990estimating} write: ``prudent ones seem to learn from the experience, not to avoid street prostitution, but to avoid being arrested... This would lead to a relatively large number of single arrests.'' A portion of individuals gain a desire or ability to avoid recapture; this portion is the amount of one-inflation in the data. The recognition of the one-inflating phenomenon has led to the development of several one-inflated distributions; the one-inflated positive Poisson \citep{godwin2017estimation}, the one-inflated zero-truncated negative binomial \citep{godwin2017one}, the one-inflated positive Poisson mixture model \citep{godwin2019one}, the one-inflated positive Poisson–Lindley \citep{tajuddin2022estimating}, and others. To our knowledge, none of these distributions have been extended to a regression framework, until now.

The development of regression models for one-inflated data entails several mathematical contributions. A novel link function for the one-inflation parameter is proposed. Marginal effects are derived. Log-likelihoods are determined. Various tests, including for no one-inflation, are presented. In addition, simulation studies verify the consistency of the proposed estimators, as well as illustrate the bias and inconsistency of the standard ZTNB estimator arising from ignoring one-inflation. A suite of tools to aid in the estimation, specification, and interpretation of the models are developed, and presented in the \texttt{R} package \texttt{oneinfl}. The importance of these contributions is highlighted through the finding of one-inflation in four applications where ZTNB has been the standard.

\section{Standard count distributions}

This section reviews some of the one-inflated distributions that have been developed to date. We start with the Poisson distribution, truncate it, and allow for one-inflation, leading to the one-inflated positive Poisson (OIPP) distribution. Allowing for unobserved heterogeneity then leads to the one-inflated zero-truncated negative binomial (OIZTNB) distribution. In the next section, we introduce regressors and build the regression models that feature in the \texttt{oneinfl} package.

\subsection{Poisson distribution}

The Poisson distribution is often the starting point for the analysis of count data, due to its relationship to more complicated but more suitable count distributions. The mass function for a random variable, \(y\), that follows a Poisson distribution, is:
\begin{equation}
f^{P} = \frac{\lambda^y}{\exp{(\lambda)}y!} \quad; \quad y = 0, 1, \dots
\end{equation}

\noindent where \(\lambda > 0\) is the mean and variance of \(y\) over the period of observation. The key limiting assumption of the Poisson model is homogeneity; the mean intensity of counts over the period of observation should be the same for all individuals in the sample. Violation of this assumption is usually dealt with by using the negative binomial model instead.


Before we develop the regression model, we discuss three main issues that cause the basic Poisson distribution to require extension. (i) Zero-truncation, the impossibility of observing a 0 count due to the administrative nature of the data. (ii) Unobserved heterogeneity, where the regressors can not explain enough of the variation in counts in order to satisfy the Poisson assumption of homogeneity. (iii) One-inflation (or deflation), a phenomenon where there is an excess (or a lack of) 1 counts.

Consideration of these issues will lead to the OIPP distribution \citep{godwin2017estimation} and the OIZTNB distribution \citep{godwin2017one}. After these models have been discussed, regressors are introduced and a regression model is developed.

\subsection{Truncation}

Only those who ``come through the door'' enter the sample, and the counts begin at 1. Such is the case with many administrative data. In general, count distributions can be altered to accommodate zero-truncation simply by recognizing that the support of the distribution is changed ($y$ cannot equal 0), and by normalizing so that the probabilities of allowable counts sum to 1:
\begin{equation}
f^{ZT} = \frac{f}{1 - f(0)} \quad ; \quad y = 1, 2, \dots
\label{eq:truncation}
\end{equation}

\noindent where \(f^{ZT}\) is the zero-truncated count distribution, \(f\) is the untruncated distribution, and $f(0)$ is the probability of a 0 under the untruncated count distribution. The zero-truncated, or positive Poisson (PP) distribution is then:
\begin{equation}
f^{PP} = \frac{\lambda^y}{(\exp{(\lambda)} - 1)y!} \quad; \quad y = 1, 2, \dots
\label{eq:PP}
\end{equation}

\noindent where \(\lambda\) is again the sole ``rate'' parameter in the distribution.

\subsection{Heterogeneity and negative binomial} \label{sec:het}

For the PP distribution in equation \ref{eq:PP} to be suitable, the Poisson assumptions of independence and homogeneity are required. Homogeneity is often too restrictive. Violation of this assumption is often detected through overdispersion of the data. A potential solution is to use a regression model, since the rigid Poisson assumption of homogeneity is more easily satisfied by allowing the Poisson parameter $\lambda$ to be a function of regressors. If there is still unobserved heterogeneity, a richer model is required. 

The negative binomial distribution is one such distribution capable of dealing with unobserved heterogeneity, and may be generated several ways. For example, one way is to include a gamma distributed error term in the function that links regressors to $\lambda$. The mass function for a random variable \(y\) that follows the negative binomial (NB) distribution is:
\begin{equation}
f^{NB} = \frac{\Gamma(\alpha + y)}{\Gamma(\alpha)\Gamma(y + 1)} \left(\frac{1}{1 + \theta}\right)^\alpha \left(\frac{\theta}{1 + \theta}\right)^y
\quad ; \quad y = 0, 1, \dots
\end{equation}

\noindent Using equation (\ref{eq:truncation}), the mass function for a random variable \(y\) that follows the zero-truncated negative binomial (ZTNB) distribution is:
\begin{equation}
f^{ZTNB} = \frac{\Gamma(\alpha + y)}{\Gamma(\alpha)\Gamma(y + 1)}\left(\frac{1}{1+\theta}\right)^\alpha \left(\frac{\theta}{1 + \theta}\right)^y \left(\frac{1}{1-(1+\theta)^{-\alpha}}\right) \quad ; \quad y = 1,2, \dots
\label{eq:ZTNB}
\end{equation}

\subsection{One-inflated count distributions}

In general, a count distribution may be 1-inflated by introducing an additional parameter $\omega$, which allows for an extra (or decreased) probability of a 1-count occurring, relative to an underlying count distribution:
\begin{equation}
\begin{split}
f^{OI} &= \omega + (1 - \omega)f(1) \quad ; \quad y = 1 \\
&= (1 - \omega)f(y) \quad; \quad y = 2, 3, \dots
\end{split}
\label{eq:OI}
\end{equation}  

\noindent where $f^{OI}$ is the one-inflated distribution, $f$ is the underlying count distribution, and $f(1)$ and $f(y)$ are probabilities of specific counts occurring. The distribution in equation \ref{eq:OI} is conceptually similar to the zero-inflated Poisson distribution \citep{johnson1969distributions,lambert1992zero}.

\subsubsection{One-inflated positive Poisson}

The one-inflated positive Poisson (OIPP) distribution \citep{godwin2017estimation} is designed to allow for one-inflation. Setting $f = f^{PP}$ in equation \ref{eq:OI} gives the mass function for the OIPP distribution:
\begin{equation}
\begin{split}
f^{OIPP} &= \omega + (1 - \omega)\frac{\lambda}{\exp{(\lambda)} - 1} \quad ; \quad y = 1 \\
&= (1 - \omega)\frac{\lambda^y}{(\exp{(\lambda)} - 1)y!} \quad; \quad y = 2, 3, \dots
\end{split}
\label{eq:OIPP}
\end{equation}    

\noindent The OIPP distribution (\ref{eq:OIPP}) deviates from the PP distribution via the parameter \(\omega\), and collapses to the PP distribution for \(\omega = 0\). The \(\lambda\) parameter has the same interpretation in both PP and OIPP distributions.

The OIPP distribution is to the PP distribution as the zero-inflated Poisson (ZIP) distribution \citep{johnson1969distributions} is to the Poisson distribution. \(\omega\) adds to the probability of a 1 count (\(\omega\) may be negative), and the probability of all counts due to the Poisson process is appropriately decreased by the scaling factor \((1 - \omega)\) so that \(f^{OIPP}\) remains a proper mass function. In order for the probability of a 1 count to remain positive, \(-\lambda(e^\lambda - \lambda - 1)^{-1} < \omega < 1\), which will be ensured in a link function for $\omega$ in Section \ref{sec:wlink}.

\subsubsection{One-inflated zero-truncated negative binomial}

As discussed in Section \ref{sec:het} on heterogeneity, the PP and OIPP distribution may not be a good choice due to their restricitive assumption of homogeneity. If there is unobserved heterogeneity then negative binomial is more appropriate than Poisson, regardless of truncation or one-inflation. The one-inflated zero-truncated negative binomial (OIZTNB) distribution \citep{godwin2017one} can be obtained by setting $f = f^{ZTNB}$ in equation \ref{eq:OI}, and has mass function:
\begin{equation}
\begin{split}
f^{OIZTNB} &= \omega + (1-\omega)\alpha\left(\frac{1}{1+\theta}\right)^\alpha\left(\frac{\theta}{1 + \theta - (1+\theta)^{1-\alpha}}\right) \ ; \ y = 1\\ 
&= (1-\omega)\frac{\Gamma(\alpha+y)}{\Gamma(\alpha)\Gamma(y+1)}\left(\frac{1}{1+\theta}\right)^\alpha\left(\frac{\theta}{1 + \theta}\right)^y \left(\frac{1}{1-(1+\theta)^{-\alpha}}\right) \ ; \ y = 2,3,\dots
\end{split}
\label{eq:OIZTNB}
\end{equation} 

\section{One-inflated count regression models}

This section presents a main contribution of the paper: the development of the one-inflated positive Poisson (OIPP) and one-inflated zero-truncated negative binomial (OIZTNB) regression models. This is accomplished by allowing the parameters of the distributions to be a function of regressors. 

\subsection{Log link for rate parameters}

In order for individual characteristics or policy interventions to influence outcomes in count data models, it is customary to link the rate parameter (denoted \(\lambda\)) to the regressors via the log link $\log(\lambda_i) = \bm{X}_i\bm{\beta}$, or:
\begin{equation}
\lambda_i = \exp(\bm{X}_i\bm{\beta)},
\label{eq:canlink}
\end{equation}

\noindent and instead estimate the parameters $\bm{\beta}$ in the link function. We follow this convention. To permit individual heterogeneity, the rate parameter has been denoted $\lambda_i$ in equation \ref{eq:canlink} (instead of $\lambda$), where the subscript $i$ denotes an individual observation, where $i = 1, \dots,n$, and where $n$ is the sample size. Equation \ref{eq:canlink} can be substituted directly into the mass function for the OIPP distribution (equation \ref{eq:OIPP}), but for the OIZTNB distribution (equation \ref{eq:OIZTNB}) an additional reparameterization is needed:
\begin{equation}
\theta = \frac{\lambda_i}{\alpha}
\label{eq:theta}
\end{equation}

\noindent Substituting the reparameterization (equation \ref{eq:theta}) and the log link (equation \ref{eq:canlink}) into the OIZTNB mass function (equation \ref{eq:OIZTNB}) gives rise to a NegBin-II type regression model (the more popular among the NegBin-I and NegBin-k type models).

\subsection{A novel link function for the inflation parameter} \label{sec:wlink}

The inflation parameter shall now be denoted \(\omega_i\) with subscript $i$ to highlight that the model allows for individual heterogeneity, and is linked to a (possibly identical to $\bm{X}_i$) set of regressors $\bm{Z}_i$. The link function for $\omega_i$ requires care and development, and does not follow any convention that we are aware of.

It is desirable for the regression models to allow for one-\textit{deflation}, as well as one-inflation. One-deflation (in a truncated model) is likely more empirically common than zero-deflation (in an un-truncated model) because of the nature of the phenomenon. While the two are conceptually and mathematically similar, zero-inflation typically occurs because ``non-users'' (always 0s) are mixed into the population of ``users''. 1-altered counts in administrative data, however, occur from a behavioural response due to the experience of being observed for the first time. The effect can be positive or negative; either the observational experience is a deterrent or an encouragement, and may vary by individual or context.

The zero-inflated Poisson model \citep{lambert1992zero} provides some guidance, where a logistic-like reparameterization is used: $\omega_i = \phi_i / (1 + \phi_i)$, and regressors are linked linearly by $\phi_i = \bm{Z}_i\bm{\gamma}$, where $\bm{\gamma}$ is an additional vector of parameters to be estimated. This ensures $\omega_i \in [0,1]$ for $\bm{Z}_i\bm{\gamma} \in (-\infty,\infty)$. If the same link were used for the one-inflated models, then $f^{OI}(1) \in [f(1),1]$\footnote{$f^{OI}$ denotes the one-inflated distribution, and $f$ denotes the underlying un-altered count distribution, as in equation \ref{eq:OI}.}. This would be acceptable if 1-deflation (a lack of 1s) is ruled out a priori, but is not an appropriate restriction for the reasons outlined above. Instead, the $\omega_i$ parameter should be bound such that the probability of a 1-count lies between 0 and 1:
\begin{equation}
f^{OI}(1) \in [0,1]
\label{eq:bindOI}
\end{equation}

\noindent That is, the inflation parameter should be able to completely alter the probability of a 1-count to any allowable value. Equation \ref{eq:bindOI} can be ensured by binding $\omega_i$ appropriately:
\begin{equation}
\omega_i \in \left[\frac{-f(1)}{1-f(1)},1\right]
\label{eq:bindomega}
\end{equation}

\noindent where the domain of $\omega_i$ is found by setting $f^{OI}(1) = 0$ and $f^{OI}(1) = 1$ in equation \ref{eq:OI}. Finally, the generalized logistic function is used to link regressors to $\omega_i$:
\begin{equation}
\omega_i = L_i + \frac{1-L_i}{1 + \exp(-\bm{Z}_i\bm{\gamma})},
\label{eq:wlink}
\end{equation}

\noindent where $L_i$ is the lower bound from equation \ref{eq:bindomega}: $L_i = \frac{-f(1)}{1-f(1)}$, $\bm{Z}_i$ is the vector of regressors affecting one-inflation, and where \(\bm{\gamma}\) is a parameter vector to be estimated. The link function in equation \ref{eq:wlink} ensures that the bounds on $\omega_i$ in equation \ref{eq:bindomega} are satisfied, and that $f^{OI}(y) \geq 0$ and $\sum f^{OI} = 1 \quad \forall \quad y = 1,2,\dots$ (provided that $f$ is a proper mass function).

For the OIPP regression model, the lower bound $L_i$ takes on the form:
\begin{equation}
L_i = \frac{-\lambda_i}{\exp(\lambda_i) - \lambda_i - 1}
\label{eq:LOIPP}
\end{equation}

\noindent and for the OIZTNB regression model the lower bound $L_i$ takes the form:
\begin{equation}
L_i=-\left\{\left(\frac{\alpha}{\alpha+\lambda_i}\right)^{-\alpha} \frac{1}{\lambda_i}\left[1+\frac{\lambda_i}{\alpha}-\left(1+\frac{\lambda_i}{\alpha}\right)^{1-\alpha}\right]-1\right\}^{-1}
\label{eq:LOIZTNB}
\end{equation}

\noindent Since $\lambda_i = \exp(\bm{X}_i \bm{\beta})$, these lower bounds $L_i$ depend on the data and vary between individuals. This must be taken into account when determining marginal effects, and when estimating the parameters of the regression models.

In summary, the OIPP regression model is obtained by substituting the log link for $\lambda_i$ (equation \ref{eq:canlink}), the logistic link for $\omega_i$, and the lower bound $L_i$ (equation \ref{eq:LOIPP}), into the OIPP mass function (equation \ref{eq:OIPP}). The OIZTNB regression model is similarly achieved, except the lower bound for $L_i$ in equation \ref{eq:LOIZTNB} must instead be used.

\section{Estimation, testing, and marginal effects}

Various issues relating to the estimation of the models are discussed in this section: maximization of the log-likelihoods; estimation of variance-covariance matrices and their related uses such as tests for no one-inflation and tests of significance; derivation and estimation of marginal effects and their standard errors; and model prediction and the visual assessment of the fitted models via plots.

\subsection{Log-likelihoods}

The \texttt{oneinfl} package estimates the OIPP and OIZTNB regression models via maximum likelihood in the \texttt{oneinfl()} function. The log-likelihoods are constructed by substituting the reparametrizations for $\lambda_i$, $L_i$, and $\theta$ (if the model is OIZTNB) into the mass function $f^{OIPP}$ or $f^{OIZTNB}$. These log-likelihoods are found in Appendix \ref{app:logl}. In addition, the \texttt{oneinfl} package also facilitates estimation of the ZTNB and PP regression models via \texttt{truncreg()}. Estimating these standard models in-house allows for easier comparison to their one-inflated counterparts, within the \texttt{oneinfl} package itself.

\subsection{Variance-covariance matrix}

The estimated variance-covariance matrix for MLEs is the negative of the expected inverse Hessian matrix, evaluated at the MLEs:
\begin{equation}
\hat{\mathbb{V}} ( \hat{\bm{\theta}} ) = -E\left[\frac{\partial^2 \ell(\hat{\bm{\theta}})}{\partial \hat{\bm{\theta}} \partial \hat{\bm{\theta}}^{\prime}}\right]^{-1}
\label{eq:varcov}
\end{equation}

\noindent The estimated standard errors are the square roots of the diagonal elements of this matrix. The Hessian matrix is not derived analytically from the log-likelihoods in equations \ref{eq:loglOIPP} or \ref{eq:logl}. Rather, the Hessian matrix is estimated numerically. The $\hat{\mathbb{V}} ( \hat{\bm{\theta}} )$ matrix is used to calculate estimated standard errors, for various tests, and for the estimated standard errors of marginal effects.

\subsection{Tests of significance}\label{sec:signifWald}

The square roots of the diagonal elements from the estimated variance-covariance matrix in equation \ref{eq:varcov} can be used to provide p-values for tests of $H_0: \beta_j = 0$ and for $H_0: \gamma_j = 0$. However, this does not typically amount to a standard test of ``significance'' in the OIPP and OIZTNB models.

The same regressors are likely to appear in both the $X$ and $Z$ matrices, allowing them to influence both the mean of the distribution through $\lambda_i$, and the amount of one-inflation through $\omega_i$. Hence, tests of significance of say $x_j$ are then joint hypotheses for when $x_j = z_m$, for example: $H_0: \beta_j = 0 \text{ and } \gamma_m = 0$. Such null hypotheses can be tested via a Wald statistic, and are accomplished via the \texttt{signifWald()} function in the \texttt{oneinfl} package.

\subsection{Tests for no one-inflation}

The amount of one-inflation can be estimated by $\frac{\sum \hat{\omega}_i}{n}$ (average one-inflation) or $\frac{\sum |\hat{\omega}_i|}{n}$ (average absolute one-inflation). Formal tests for the presence of one-inflation are also developed. Recalling equation \ref{eq:wlink}, a test for no one-inflation amounts to $H_0:\bm{\gamma} = \bm{0}$, rejection of which would indicate that the one-inflated model is more appropriate. Alternatively, the null and alternative hypotheses could be specified as, for example, $H_0:\text{ZTNB}$ and $H_A:\text{OIZTNB}$. The null hypothesis is nested in the alternative, allowing for the asymptotically equivalent Wald, likelihood ratio, and Lagrange multiplier testing. Only the Wald and likelihood ratio tests are developed here.

\subsubsection{Wald test}

The Wald test statistic does not require estimation of the model under the null hypothesis, and is calculated by:
\begin{equation}
W = \bm{\gamma}^{\prime} \left[R \hat{\mathbb{V}} ( \hat{\bm{\theta}} ) R^{\prime}\right]^{-1} \bm{\gamma}
\end{equation}

\noindent where the $R$ matrix selects the necessary rows and columns from $\hat{\mathbb{V}}(\hat{\theta})$. This Wald statistic is $\chi^2_{(p)}$ distributed, where $p$ are the number of regressors in $Z$ including the intercept. Note that the null hypothesis is not on the boundary of the parameter space as one-deflation is allowed in the models through the choice of the link function in equations \ref{eq:bindOI} and \ref{eq:bindomega}, otherwise the Wald statistic would follow a half Chi-square distribution. The Wald test statistic and associated p-value for no one-inflation are calculated by the \texttt{oneWald()} function in the \texttt{oneinfl} package.

\subsubsection{Likelihood ratio test}

The likelihood ratio test statistic for testing the null of no one-inflation can be obtained by comparing the log-likelihood functions (evaluated at the MLEs) of OIPP and PP, or OIZTNB and ZTNB. The statistic takes the form $LRT = -2(\hat{\ell}^{PP} - \hat{\ell}^{OIPP})$ or $LRT = -2(\hat{\ell}^{ZTNB} - \hat{\ell}^{OIZTNB})$ and is $\chi^2_{(p)}$, where $p$ is again the rows in $\boldmath{\gamma}$ (also the columns of $Z$). The \texttt{oneLRT()} function in the \texttt{oneinfl} package provides this test statistic and associated p-value by evaluating both a one-inflated model (OIPP or OIZTNB) and standard counterpart (PP or ZTNB).

\subsection{Estimated marginal effects and their standard errors}

Of particular importance are the marginal effects that individual characteristics or policies in $X$ and $Z$ have on the expected number of counts $\mathbb{E}[y_i]$. Obtaining the estimated marginal effects is a standard goal of regression modelling. The following derivatives need to be determined:
\begin{equation}
\frac{\partial \mathbb{E}[y_i]} {\partial q_{ij}},
\nonumber
\end{equation}

\noindent where $q_{ij}$ is the $j^{th}$ regressor in \(X\) and/or \(Z\), for individual $i$. These partial derivatives are derived in Appendix \ref{app:marginal}. The marginal effects in equations \ref{eq:marginalOIPP} or \ref{eq:marginalOIZTNB} are evaluated at the MLEs, and either averaged over all data points (``average effects''), evaluated once at the sample means of the data (``effect at means''), or at some representative case $\boldmath{X}_i, \boldmath{Z}_i$. For binary dummy variables, the ``marginal effect'' is instead $\mathbb{E}[y_i \mid D_{ij} = 1] - \mathbb{E}[y_i \mid D_{ij} = 0]$, for some dummy in \(X\) and/or \(Z\). This is straightforward to estimate by evaluating the expected difference at multiple data points.

Model parameters from PP or ZTNB models are incident risk ratios when exponentiated, offering an interpretation of the impact of a regressor on the outcome variable. However, this is not possible for the OIPP and OIZTNB models, since each regressor typically enters the model through a mean link (equation \ref{eq:canlink}) \textit{and} through a one-inflation link (equation \ref{eq:wlink}). Marginal effects offer an alternative way to compare the impact of regressors between ZTNB and OIZTNB.

In order to determine whether the estimated marginal effects are significant, their standard errors must be estimated. There are two widely popular options available; the delta method and the bootstrap. We employ the delta method since significant computational time can be required in order to estimate some one-inflated models.

The delta method can provide the variance of a function \(h(\boldmath{\hat{\theta}})\), where \(\boldmath{\hat{\theta}}\) has a known asymptotic distribution (Normal in the case of MLEs) with variance-covariance matrix \(\mathbb{V}(\boldmath{\hat{\theta}})\). Justified by a first order Taylor series approximation, the asymptotic variance of \(h(\boldmath{\hat{\theta}})\) is approximately equal to $J(\hat{\boldmath{\theta}})\hat{\mathbb{V}}(\hat{\theta})J(\hat{\boldmath{\theta}})^{\prime}$. The Jacobian matrix $J(\hat{\boldmath{\theta}})$ is the vector of first partial derivatives of \(h(\boldmath{\hat{\theta}})\). Standard errors for the marginal effects are estimated by setting \(h(\boldmath{\hat{\theta}})\) equal to the marginal effects derived in equations \ref{eq:marginalOIPP} or \ref{eq:marginalOIZTNB}, numerically approximating $J(\hat{\boldmath{\theta}})$ using the \texttt{numericDeriv()} function in base \texttt{R} \citep{R}, and taking the square roots of the diagonal elements of $J(\hat{\boldmath{\theta}})\hat{\mathbb{V}}(\hat{\theta})J(\hat{\boldmath{\theta}})^{\prime}$. These standard errors, as well as \textit{z}-tests for the significance of the marginal effects, are reported in the \texttt{margins()} function in the \texttt{oneinfl} package.

\subsection{Predicted counts and visualizing the fit of the models}

A model's predicted counts are obtained by evaluating its probability mass function (such as equation \ref{eq:OIZTNB}) at each $\hat{\lambda}_i$ (and $\hat{\omega}_i$ if relevant), and summing. This is facilitated in the \texttt{oneinfl} package using the \texttt{pred()} function.

Visualizing the data and predicted counts is an important diagnostic tool, as often the presence of one-inflation is immediately apparent. The \texttt{oneplot()} function in the \texttt{oneinfl} package generates a barplot of the actual data, as well as plotting the predicted counts for up to four models: positive Poisson (PP), one-inflated positive Poisson (OIPP), zero truncated negative binomial (ZTNB), and one-inflated zero truncated negative binomial (OIZTNB). Figures \ref{fig:medpar} - \ref{fig:Alaska} show the result of applying \texttt{oneplot()} to ZTNB and OIZTNB models.

\section{Simulation studies}

{\linespread{1}
\begin{table}[h]
\begin{center}
\begin{tabular}{ccccccccc}
model                       & $n$                                        & $\hat{\beta}_0$ & $\hat{\beta}_1$ & $\hat{\beta}_2$ & $\hat{\gamma}_0$ & $\hat{\gamma}_1$ & $\hat{\gamma}_2$ & $\hat{\alpha}$ \\ \hline
\multicolumn{9}{c}{Poisson}                                                                                                                                                                              \\ \hline
\multicolumn{1}{c|}{OIPP}   & \multicolumn{1}{c|}{\multirow{2}{*}{200}}  & 0.17            & 0.05            & -0.24           & 3.24             & 3.28             & 3.02             &                \\
\multicolumn{1}{c|}{PP}     & \multicolumn{1}{c|}{}                      & -258.08         & -141.20         & -75.62          &                  &                  &                  &                \\ \hline
\multicolumn{1}{c|}{OIPP}   & \multicolumn{1}{c|}{\multirow{2}{*}{400}}  & -0.18           & -0.12           & 0.00            & 1.29             & 1.29             & 1.52             &                \\
\multicolumn{1}{c|}{PP}     & \multicolumn{1}{c|}{}                      & -257.03         & -140.63         & -74.45          &                  &                  &                  &                \\ \hline
\multicolumn{1}{c|}{OIPP}   & \multicolumn{1}{c|}{\multirow{2}{*}{800}}  & -0.01           & -0.02           & 0.22            & 0.69             & 0.69             & 0.73             &                \\
\multicolumn{1}{c|}{PP}     & \multicolumn{1}{c|}{}                      & -256.58         & -140.36         & -74.43          &                  &                  &                  &                \\ \hline
\multicolumn{1}{c|}{OIPP}   & \multicolumn{1}{c|}{\multirow{2}{*}{1600}} & -0.19           & -0.11           & 0.18            & 0.34             & 0.34             & -0.08            &                \\
\multicolumn{1}{c|}{PP}     & \multicolumn{1}{c|}{}                      & -256.46         & -140.29         & -74.04          &                  &                  &                  &                \\ \hline
\multicolumn{9}{c}{Negative Binomial}                                                                                                                                                                    \\ \hline
\multicolumn{1}{c|}{OIZTNB} & \multicolumn{1}{c|}{\multirow{2}{*}{200}}  & -0.10           & -0.10           & -0.67           & 3.22             & 3.25             & 2.99             & 27.77          \\
\multicolumn{1}{c|}{ZTNB}   & \multicolumn{1}{c|}{}                      & -331.59         & -194.82         & -89.34          &                  &                  &                  & -96.02         \\ \hline
\multicolumn{1}{c|}{OIZTNB} & \multicolumn{1}{c|}{\multirow{2}{*}{400}}  & -0.38           & -0.23           & -0.08           & 1.27             & 1.28             & 1.50             & 11.26          \\ \cline{1-1}
\multicolumn{1}{c|}{ZTNB}   & \multicolumn{1}{c|}{}                      & -325.28         & -190.76         & -85.91          &                  &                  &                  & -96.20         \\ \hline
\multicolumn{1}{c|}{OIZTNB} & \multicolumn{1}{c|}{\multirow{2}{*}{800}}  & -0.08           & -0.06           & 0.22            & 0.69             & 0.69             & 0.72             & 5.42           \\ \cline{1-1}
\multicolumn{1}{c|}{ZTNB}   & \multicolumn{1}{c|}{}                      & -321.50         & -188.46         & -85.61          &                  &                  &                  & -96.28         \\ \hline
\multicolumn{1}{c|}{OIZTNB} & \multicolumn{1}{c|}{\multirow{2}{*}{1600}} & -0.11           & -0.07           & 0.27            & 0.32             & 0.33             & -0.14            & 2.44           \\ \cline{1-1}
\multicolumn{1}{c|}{ZTNB}   & \multicolumn{1}{c|}{}                      & -320.42         & -187.74         & -84.64          &                  &                  &                  & -96.32         \\ \hline
\end{tabular}
\end{center}
\caption{Percent biases of the MLEs from various models, under a one-inflated DGP (either OIPP or OIZTNB).}
\label{tab:MC}
\end{table}
}

Simulation studies are undertaken to examine the percent bias of various maximum likelihood estimators (MLEs) under different data generating processes (DGPs) and for differently specified models. The simulations suggest massive bias of the MLEs for when one-inflation is present, but is ignored. That is, the MLEs from the standard positive Poisson (PP) or zero-truncated negative binomial (ZTNB) exhibit large bias when the true data generating process is characteristic of one-inflation. The bias of the estimators depends on the values of the parameters chosen in the DGP, but it is not difficult to find values for which there is over 100\% bias. 

We conduct two main simulations, one each for a one-inflated positive Poisson (OIPP) DGP and for a one-inflated zero-truncated negative binomial (OIZTNB) DGP. Under each DGP, both the standard and one-inflated models are estimated. The same two regressors are used in the $X$ and $Z$ matrices; $x_1$ is generated according to a $N(10,1)$ distribution and $x_2$ is a dummy variable with mean 0.5. The parameter values under OIPP are chosen to be $\boldmath{\beta} = (-2, 0.4, 0.2)$, $\boldmath{\gamma} = (-21, 2, 0.5)$. The same parameter values are chosen for OIZTNB, but with the addition of $\alpha = 10$. The simulations are run for various sample sizes, and all experiments use 10,000 replications. Random variates for OIPP and OIZTNB were generated using the \texttt{roipp()} and \texttt{roiztnb()} functions in \texttt{oneinfl}. The results of the experiments are reported in Table \ref{tab:MC}.

Table \ref{tab:MC} shows very high percent biases for PP and ZTNB (between 75\% and 325\% bias), that do not diminish as $n$ increases, reflecting the inconsistency of the estimators in the presence of one-inflation. In contrast, the percent biases of OIPP and OIZTNB are always less than 4\%, and diminish further as $n$ increases. These results reflect the consistency of a correctly specified MLE.

\section{Illustrations}

Four empirical illustration are examined. All data have traditionally been used to exhibit the zero-truncated negative binomial (ZTNB) model; in all cases one-inflation is found and the one-inflated zero-truncated negative binomial (OIZTNB) model is superior. The illustrations highlight how the estimated parameters and interpretations of the models change, sometimes dramatically, with the incorporation of one-inflation.

\subsection{Hospital length of stay: MedPar data} \label{sec:medpar}

\begin{figure}
\centering
\includegraphics[scale=1]{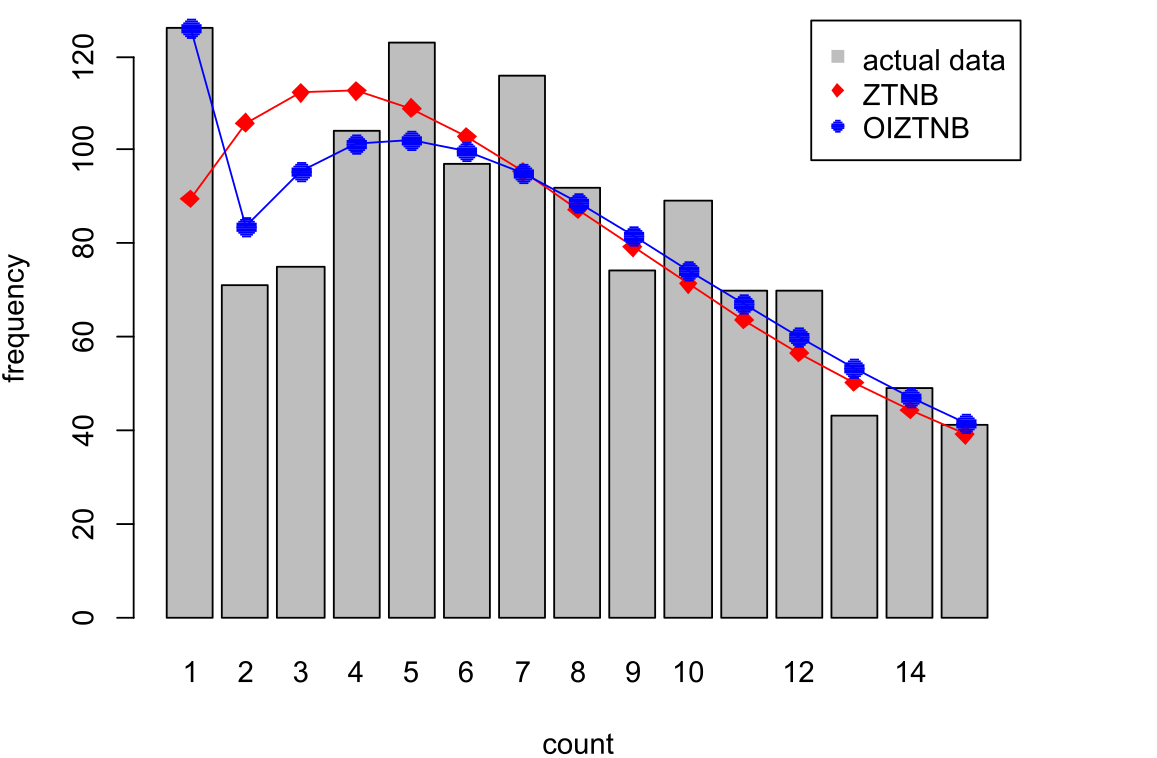}
\caption{The one-inflated zero truncated negative binomial (OIZTNB) regression model, and the zero truncated negative binomial (ZTNB) regression model, fit to the MedPar data set.}
\label{fig:medpar}
\end{figure}

{\linespread{1}
\begin{table}[h]
\vspace{1mm}
\centering
\begin{tabular}{cc|cc|cc}
\hline
                                 & ZTNB               & \multicolumn{2}{c|}{OIZTNB}                                   & \multicolumn{1}{c|}{ZTNB}           & OIZTNB         \\ \hline
                                 & $\hat{\boldmath{\beta}}$ & \multicolumn{1}{c|}{$\hat{\boldmath{\beta}}$} & $\hat{\boldmath{\gamma}}$ & \multicolumn{2}{c}{marginal effects}                 \\ \hline
\multicolumn{1}{c|}{(Intercept)} & 2.333$^{***}$      & \multicolumn{1}{c|}{2.299$^{***}$}      & -4.200$^{***}$      &                                     &                \\
\multicolumn{1}{c|}{}            & (0.075)            & \multicolumn{1}{c|}{(0.072)}            & (0.506)             &                                     &                \\ \hline
\multicolumn{1}{c|}{white}       & -0.132$^{*}$       & \multicolumn{1}{c|}{-0.097}             & 0.659               & \multicolumn{1}{c|}{-1.296$^{*}$}   & -1.258$^{*}$   \\
\multicolumn{1}{c|}{}            & (0.075)            & \multicolumn{1}{c|}{(0.071)}            & (0.481)             & \multicolumn{1}{c|}{(0.776)}        & (0.734)        \\ \hline
\multicolumn{1}{c|}{died}        & -0.251$^{***}$     & \multicolumn{1}{c|}{-0.068}             & 2.335$^{***}$       & \multicolumn{1}{c|}{-2.238$^{***}$} & -2.189$^{***}$ \\
\multicolumn{1}{c|}{}            & (0.045)            & \multicolumn{1}{c|}{(0.045)}            & (0.236)             & \multicolumn{1}{c|}{(0.387)}        & (0.396)        \\ \hline
\multicolumn{1}{c|}{type2}       & 0.260$^{***}$      & \multicolumn{1}{c|}{0.234$^{***}$}      & -0.541$^{*}$        & \multicolumn{1}{c|}{2.639$^{***}$}  & 2.575$^{***}$  \\
\multicolumn{1}{c|}{}            & (0.055)            & \multicolumn{1}{c|}{(0.054)}            & (0.283)             & \multicolumn{1}{c|}{(0.616)}        & (0.588)        \\ \hline
\multicolumn{1}{c|}{type3}       & 0.769$^{***}$      & \multicolumn{1}{c|}{0.756$^{***}$}      & -0.751$^{*}$        & \multicolumn{1}{c|}{10.159$^{***}$} & 10.142$^{***}$ \\
\multicolumn{1}{c|}{}            & (0.083)            & \multicolumn{1}{c|}{(0.079)}            & (0.447)             & \multicolumn{1}{c|}{(1.530)}        & (1.467)        \\ \hline
\multicolumn{6}{l}{Significance at the 1\% (***), 5\% (**), and 10\% (*) levels.}                                                                                                                                       
\end{tabular}
\caption{ZTNB and OIZTNB models fit to the MedPar data.}
\label{tab:medpar}
\end{table}
}

A particular data set from the 1991 Arizona MedPar database has championed and exemplified the zero-truncated negative binomial (ZTNB) regression model. The data appears in articles \citep{hardin2007generalized, hardin2015regression}, books \citep{hilbe2011negative, hilbe2014modeling}, R packages \citep{COUNT, msme}, and several web vignettes for ZTNB estimation. ZTNB regression models have been estimated for subsequent waves of MedPar data in many research articles \citep{baser2010impact, kleindorfer2009us, lasater2021evaluation}, to name a few.

The goal is to model a patient's hospital length of stay (\texttt{los}), which is a zero-truncated count. Some illustrations of the ZTNB model use membership to a health maintenance organization as an explanatory variable, but we instead estimate the model from ``11.1: Zero-truncated count models'', in \cite{hilbe2011negative}. All of the explanatory variables are dummies: \texttt{white == 1} if the patient identifies as Caucasian; \texttt{died == 1} if the patient died during their stay; and categorical dummies for the type of admission, whether elective, urgent (\texttt{type2 == 1}) or emergency (\texttt{type3 == 1}). 

The estimated ZTNB and OIZTNB models, and the marginal effects from each, are displayed in Table \ref{tab:medpar}. Average and absolute one-inflation is 4.2\% and 6.8\% respectively. One-inflation is plausible in such data, for at least two reasons. Suppose there are two types of individuals mixed into the sample; ill and healthy. Type is revealed through examination by a doctor, when \texttt{los = 1}. Those that are healthy are discharged and one-inflate the data. As a second channel for one-inflation, suppose that patients that die tend to do so on their first day of stay. Both Wald and LRT tests of no one-inflation support the presence of one-inflation with test statistics of 469.1 and 131.4 respectively, giving associated p-values of 0.

Despite its association with the MedPar data set, the ZTNB regression model is likely misspecified, and the MLEs are biased and inconsistent.

A plot of the actual data, as well as the predicted counts from the ZTNB and OIZTNB models, is generated using \texttt{oneplot()} and displayed in Figure \ref{fig:medpar}. One-inflation is visibly apparent. 

All estimates and standard errors for the parameters and marginal effects change somewhat between ZTNB and OIZTNB models. Most notably is for that of the regressor \texttt{died}, whose association to \texttt{los} occurs through one-inflation rather than through the mean of the underlying count distribution. In addition, under OIZTNB, \texttt{white} is no longer a significant variable according to the Wald test from Section \ref{sec:signifWald}, which provides a test statistics of 3.73 and associated p-value of 0.16.

\subsection{Hospital length of stay: Arizona Medicare data}

\begin{figure}[!t]
\centering
\includegraphics[scale=1]{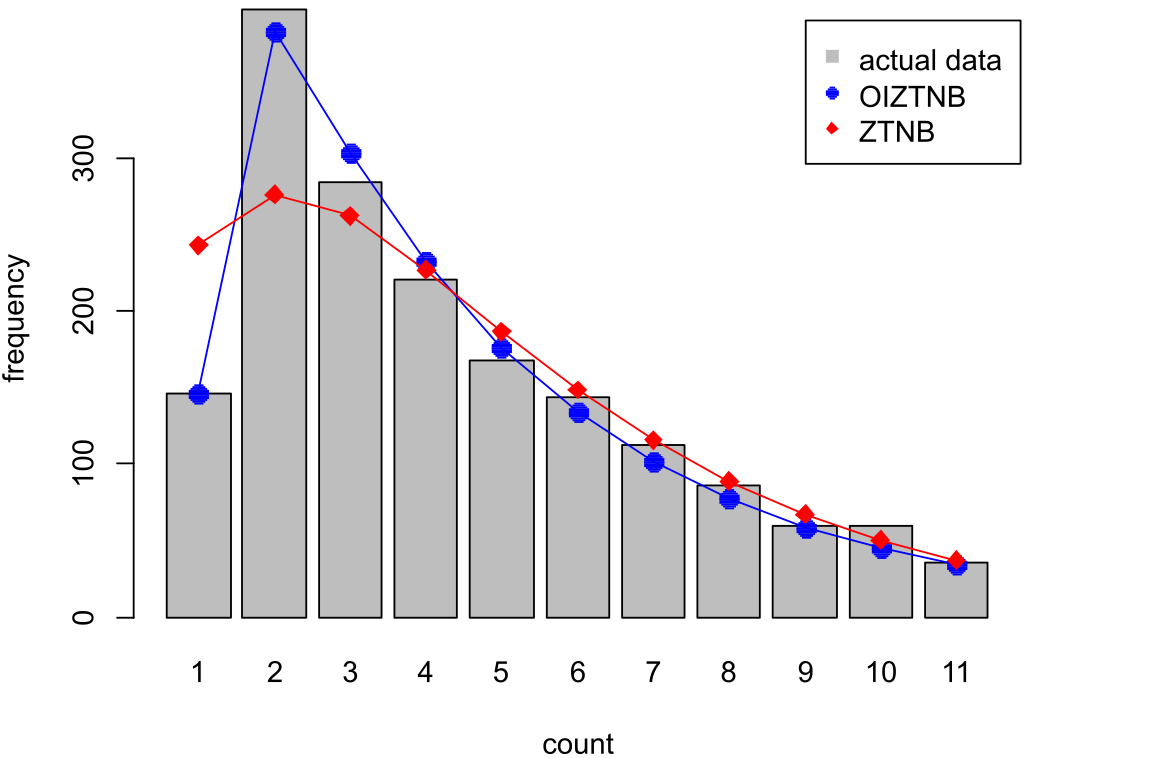}
\caption{OIZTNB and ZTNB regression models fit to the Arizona Medicare data set.}
\label{fig:azdrg112}
\end{figure}

{\linespread{1}
\begin{table}[!b]
\centering
\begin{tabular}{cccccc}
\hline
                                   & \multicolumn{1}{c|}{ZTNB}               & \multicolumn{2}{c|}{OIZTNB}                                                        & \multicolumn{1}{c|}{ZTNB}           & OIZTNB         \\ \hline
                                   & \multicolumn{1}{c|}{$\hat{\boldmath{\beta}}$} & \multicolumn{1}{c|}{$\hat{\boldmath{\beta}}$} & \multicolumn{1}{c|}{$\hat{\boldmath{\gamma}}$} & \multicolumn{2}{c}{marginal effects}                 \\ \hline
\multicolumn{1}{c|}{(Intercept)} & \multicolumn{1}{c|}{1.051$^{***}$}      & \multicolumn{1}{c|}{0.744$^{***}$}      & \multicolumn{1}{c|}{-2.131$^{***}$}      &                                     &                \\
\multicolumn{1}{c|}{}              & \multicolumn{1}{c|}{(0.043)}            & \multicolumn{1}{c|}{(0.066)}            & \multicolumn{1}{c|}{(0.199)}             &                                     &                \\ \hline
\multicolumn{1}{c|}{gender}      & \multicolumn{1}{c|}{-0.166$^{***}$}     & \multicolumn{1}{c|}{-0.178$^{***}$}     & \multicolumn{1}{c|}{0.597$^{***}$}        & \multicolumn{1}{c|}{-0.729$^{***}$} & -0.727$^{***}$ \\
\multicolumn{1}{c|}{}              & \multicolumn{1}{c|}{(0.035)}            & \multicolumn{1}{c|}{(0.047)}            & \multicolumn{1}{c|}{(0.205)}             & \multicolumn{1}{c|}{(0.160)}        & (0.165)        \\ \hline
\multicolumn{1}{c|}{type1}       & \multicolumn{1}{c|}{0.737$^{***}$}      & \multicolumn{1}{c|}{0.922$^{***}$}      & \multicolumn{1}{c|}{-1.340$^{***}$}      & \multicolumn{1}{c|}{2.699$^{***}$}  & 2.697$^{***}$  \\
\multicolumn{1}{c|}{}              & \multicolumn{1}{c|}{(0.040)}            & \multicolumn{1}{c|}{(0.056)}            & \multicolumn{1}{c|}{(0.183)}             & \multicolumn{1}{c|}{(0.134)}        & (0.136)        \\ \hline
\multicolumn{1}{c|}{age75}       & \multicolumn{1}{c|}{0.131$^{***}$}      & \multicolumn{1}{c|}{0.155$^{***}$}       & \multicolumn{1}{c|}{-0.123}              & \multicolumn{1}{c|}{0.560$^{***}$}  & 0.552$^{***}$   \\
\multicolumn{1}{c|}{}              & \multicolumn{1}{c|}{(0.038)}            & \multicolumn{1}{c|}{(0.050)}            & \multicolumn{1}{c|}{(0.206)}             & \multicolumn{1}{c|}{(0.173)}        & (0.181)        \\ \hline
\multicolumn{6}{l}{Significance at the 1\% (***), 5\% (**), and 10\% (*) levels.}                                                                                                                                       
\end{tabular}
\caption{ZTNB and OIZTNB models fit to the Arizona Medicare data.}
\label{tab:azdrg112}
\end{table}
}

This illustration uses 1995 Arizona Medicare data for the hospital length of stay for patients undergoing a heart procedure. The data may be found in the \texttt{R} package \texttt{COUNT} \citep{COUNT}. This data has appeared as an application of the ZTNB model, and a two component truncated negative binomial mixture model, in \cite{hilbe2007negative, hilbe2011negative}. Note that finite mixture models are incapable of capturing one-inflation, since the mixture component would need to reach the boundary (i.e. $\lambda \to 0$). Regressors are again all dummies: \texttt{gender == 1} for male; \texttt{type1 == 1} for emergency/urgent admission and \texttt{== 0} for elective admission; and \texttt{age75 == 1} if the patient is over the age of 75.

OIZTNB and ZTNB models and marginal effects are estimated similar to Section \ref{sec:medpar}. Results are reported in Table \ref{tab:azdrg112}. Actual and predicted counts appear in Figure \ref{fig:azdrg112}. One-deflation is present (average one-inflation is estimated to be -16.4\%), and is seemingly driven by type of admission. A one-count is less probable under an emergency/urgent admission. Wald and LRT tests for no one-inflation again produce \textit{p}-values of 0. Table \ref{tab:azdrg112} shows big swings in the values of the estimated $\hat{\boldmath{\beta}}$ between ZTNB and OIZTNB, as well as some moderate changes in the estimated marginal effects.

\subsection{Recreation demand}

\begin{figure}[!t]
\centering
\includegraphics[scale=1]{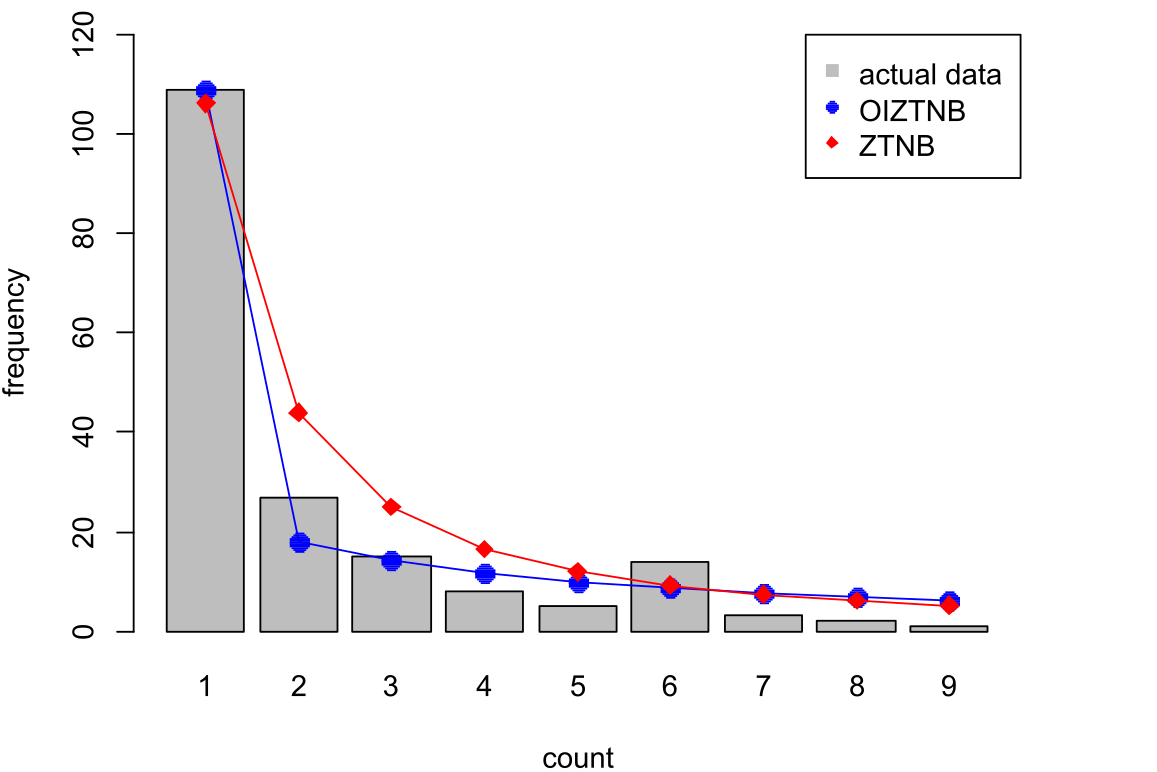}
\caption{OIZTNB and ZTNB regression models fit to the \cite{loomis2003travel} data set.}
\label{fig:Loomis}
\end{figure}

{\linespread{1}
\begin{table}[!b]
\centering
\begin{tabular}{cccccc}
\hline
                                 & \multicolumn{1}{c|}{ZTNB}               & \multicolumn{2}{c|}{OIZTNB}                                                        & \multicolumn{1}{c|}{ZTNB}           & OIZTNB         \\ \hline
                                 & \multicolumn{1}{c|}{$\hat{\boldmath{\beta}}$} & \multicolumn{1}{c|}{$\hat{\boldmath{\beta}}$} & \multicolumn{1}{c|}{$\hat{\boldmath{\gamma}}$} & \multicolumn{2}{c}{marginal effects}                 \\ \hline
\multicolumn{1}{c|}{(Intercept)} & \multicolumn{1}{c|}{-0.326}             & \multicolumn{1}{c|}{3.415$^{***}$}      & \multicolumn{1}{c|}{-3.480$^{***}$}      &                                     &                \\
\multicolumn{1}{c|}{}            & \multicolumn{1}{c|}{(1.024)}            & \multicolumn{1}{c|}{(0.430)}            & \multicolumn{1}{c|}{(0.986)}             &                                     &                \\ \hline
\multicolumn{1}{c|}{gender}      & \multicolumn{1}{c|}{0.841$^{***}$}      & \multicolumn{1}{c|}{0.469$^{**}$}       & \multicolumn{1}{c|}{-0.531}              & \multicolumn{1}{c|}{14.01$^{***}$}  & 8.55$^{***}$   \\
\multicolumn{1}{c|}{}            & \multicolumn{1}{c|}{(0.281)}            & \multicolumn{1}{c|}{(0.196)}            & \multicolumn{1}{c|}{(0.343)}             & \multicolumn{1}{c|}{(3.55)}         & (1.99)         \\ \hline
\multicolumn{1}{c|}{income2}     & \multicolumn{1}{c|}{1.166$^{**}$}       & \multicolumn{1}{c|}{0.189}              & \multicolumn{1}{c|}{0.347}               & \multicolumn{1}{c|}{41.26$^{*}$}    & 4.154          \\
\multicolumn{1}{c|}{}            & \multicolumn{1}{c|}{(0.484)}            & \multicolumn{1}{c|}{(0.285)}            & \multicolumn{1}{c|}{(0.604)}             & \multicolumn{1}{c|}{(23.85)}        & (7.76)         \\ \hline
\multicolumn{1}{c|}{income3}     & \multicolumn{1}{c|}{-0.958$^{**}$}      & \multicolumn{1}{c|}{-0.526}             & \multicolumn{1}{c|}{1.142$^{*}$}         & \multicolumn{1}{c|}{-24.40$^{**}$}  & -14.20$^{**}$  \\
\multicolumn{1}{c|}{}            & \multicolumn{1}{c|}{(0.416)}            & \multicolumn{1}{c|}{(0.313)}            & \multicolumn{1}{c|}{(0.588)}             & \multicolumn{1}{c|}{(9.56)}         & (6.08)         \\ \hline
\multicolumn{1}{c|}{income4}     & \multicolumn{1}{c|}{-0.326$^{**}$}      & \multicolumn{1}{c|}{-0.780$^{*}$}       & \multicolumn{1}{c|}{0.003}               & \multicolumn{1}{c|}{-10.28}         & -17.00$^{***}$ \\
\multicolumn{1}{c|}{}            & \multicolumn{1}{c|}{(0.430)}            & \multicolumn{1}{c|}{(0.270)}            & \multicolumn{1}{c|}{(0.555)}             & \multicolumn{1}{c|}{(12.19)}        & (5.66)         \\ \hline
\multicolumn{1}{c|}{travel2}     & \multicolumn{1}{c|}{-0.582}             & \multicolumn{1}{c|}{-0.499$^{***}$}     & \multicolumn{1}{c|}{1.983$^{***}$}       & \multicolumn{1}{c|}{-20.31}         & -16.78$^{***}$ \\
\multicolumn{1}{c|}{}            & \multicolumn{1}{c|}{(0.358)}            & \multicolumn{1}{c|}{(0.209)}            & \multicolumn{1}{c|}{(0.767)}             & \multicolumn{1}{c|}{(13.97)}        & (5.99)         \\ \hline
\multicolumn{1}{c|}{travel3}     & \multicolumn{1}{c|}{-4.141$^{***}$}     & \multicolumn{1}{c|}{-2.142$^{***}$}     & \multicolumn{1}{c|}{4.828$^{***}$}       & \multicolumn{1}{c|}{-54.93$^{***}$} & -40.75$^{***}$ \\
\multicolumn{1}{c|}{}            & \multicolumn{1}{c|}{(0.356)}            & \multicolumn{1}{c|}{(0.289)}            & \multicolumn{1}{c|}{(0.763)}             & \multicolumn{1}{c|}{(12.83)}        & (4.59)         \\ \hline
\multicolumn{6}{l}{Significance at the 1\% (***), 5\% (**), and 10\% (*) levels.}                                                                                                                                     
\end{tabular}
\caption{ZTNB and OIZTNB models fit to the \cite{loomis2003travel} data.}
\label{tab:Loomis}
\end{table}
}

The data is from \cite{loomis2003travel} and is on number of trips taken to a recreation site. Estimates of the effect of travel cost on the number of visits can be used to value the site. Endogenous stratification is of concern for this type of data; it is collected on-site, and risks over-sampling avid users. However, in this illustration, and in other examples for recreation demand, endogenous stratification seems to have little impact. For this data, \cite{hilbe2011negative} finds that the ``zero-truncated model without adjustment for endogenous stratification is preferred.'' Additionally, he writes, ``the extreme number of 1s in the data is the likely cause of the model not fitting.'' Descriptions of the variables can be found in \cite{loomis2003travel} or the \texttt{R} package \texttt{COUNT} \citep{COUNT}.

ZTNB and OIZTNB regression models are estimated for the data (see Figure \ref{fig:Loomis}). Massive changes in the estimated $\hat{\boldmath{\beta}}$ and the marginal effects can be seen in Table \ref{tab:Loomis}, namely the effect of \texttt{income2} changes from 41.26 to 4.154. Average one-inflation is estimated to be 24\% (with some observations having one-deflation), and both Wald and LRT tests (\textit{p}-values 0) again support the presence of one-inflation.

\subsection{Fishing trips}

\begin{figure}[!t]
\centering
\includegraphics[scale=1]{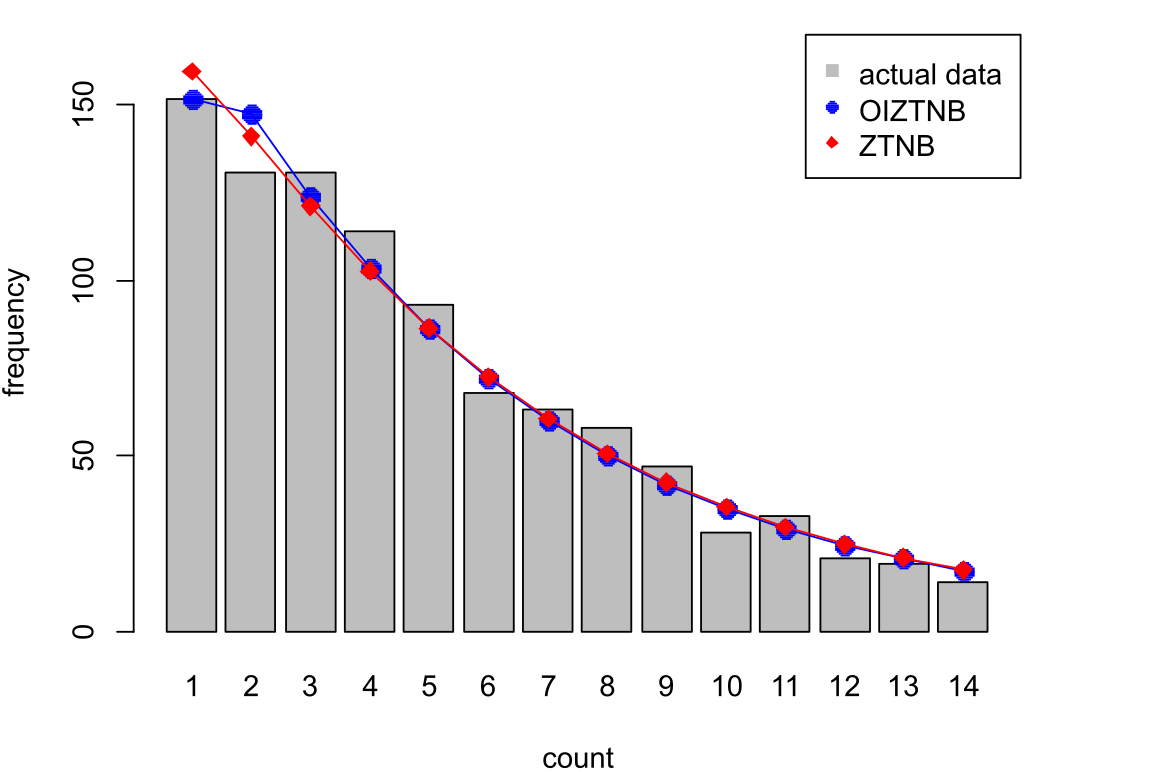}
\caption{OIZTNB and ZTNB regression models fit to the fishing trip data set.}
\label{fig:Alaska}
\end{figure}

{\linespread{1}
\begin{table}[!t]
\centering
\begin{tabular}{cccccc}
\hline
                                 & \multicolumn{1}{c|}{ZTNB}               & \multicolumn{2}{c|}{OIZTNB}                                                                     & \multicolumn{1}{c|}{ZTNB}           & OIZTNB         \\ \hline
                                 & \multicolumn{1}{c|}{$\hat{\boldmath{\beta}}$} & \multicolumn{1}{c|}{$\hat{\boldmath{\beta}}$}              & \multicolumn{1}{c|}{$\hat{\boldmath{\gamma}}$} & \multicolumn{2}{c}{marginal effects}                 \\ \hline
\multicolumn{1}{c|}{(Intercept)} & \multicolumn{1}{c|}{2.152$^{***}$}      & \multicolumn{1}{c|}{2.188$^{***}$}                   & \multicolumn{1}{c|}{-2.322$^{***}$}      &                                     &                \\
\multicolumn{1}{c|}{}            & \multicolumn{1}{c|}{(0.081)}            & \multicolumn{1}{c|}{(0.103)}                         & \multicolumn{1}{c|}{(0.284)}             &                                     &                \\ \hline
\multicolumn{1}{c|}{AVLONG}      & \multicolumn{1}{c|}{-0.257$^{***}$}     & \multicolumn{1}{c|}{-0.284$^{***}$}                  & \multicolumn{1}{c|}{0.255$^{***}$}       & \multicolumn{1}{c|}{1.446$^{***}$}  & -1.674$^{***}$ \\
\multicolumn{1}{c|}{}            & \multicolumn{1}{c|}{(0.022)}            & \multicolumn{1}{c|}{(0.027)}                         & \multicolumn{1}{c|}{(0.046)}             & \multicolumn{1}{c|}{(0.139)}        & (0.166)        \\ \hline
\multicolumn{1}{c|}{MISS}        & \multicolumn{1}{c|}{-0.151$^{**}$}      & \multicolumn{1}{c|}{-0.147$^{*}$}                    & \multicolumn{1}{c|}{0.191}               & \multicolumn{1}{c|}{-0.814$^{**}$}  & -0.795$^{**}$  \\
\multicolumn{1}{c|}{}            & \multicolumn{1}{c|}{(0.070)}            & \multicolumn{1}{c|}{(0.078)}                         & \multicolumn{1}{c|}{(0.206)}             & \multicolumn{1}{c|}{(0.364)}        & (0.366)        \\ \hline
\multicolumn{1}{c|}{CROWD}       & \multicolumn{1}{c|}{-0.034$^{***}$}     & \multicolumn{1}{c|}{-0.026}                          & \multicolumn{1}{c|}{0.075}               & \multicolumn{1}{c|}{-0.192}         & -0.176         \\
\multicolumn{1}{c|}{}            & \multicolumn{1}{c|}{(0.030)}            & \multicolumn{1}{c|}{(0.033)}                         & \multicolumn{1}{c|}{(0.091)}             & \multicolumn{1}{c|}{(0.168)}        & (0.186)        \\ \hline
\multicolumn{1}{c|}{INC}         & \multicolumn{1}{c|}{0.000}              & \multicolumn{1}{c|}{0.001}                           & \multicolumn{1}{c|}{0.002}               & \multicolumn{1}{c|}{0.002}          & 0.004          \\
\multicolumn{1}{c|}{}            & \multicolumn{1}{c|}{(0.001)}            & \multicolumn{1}{c|}{(0.001)}                         & \multicolumn{1}{c|}{(0.003)}             & \multicolumn{1}{c|}{(0.005)}        & (0.006)        \\ \hline
\multicolumn{1}{c|}{CPMILE}      & \multicolumn{1}{c|}{1.188$^{***}$}      & \multicolumn{1}{c|}{1.013$^{**}$} & \multicolumn{1}{c|}{-3.028$^{**}$}       & \multicolumn{1}{c|}{6.685$^{***}$}  & 6.941$^{***}$  \\
\multicolumn{1}{c|}{}            & \multicolumn{1}{c|}{(0.407)}            & \multicolumn{1}{c|}{(0.443)}                         & \multicolumn{1}{c|}{(1.462)}             & \multicolumn{1}{c|}{(2.317)}        & (2.532)        \\ \hline
\multicolumn{1}{c|}{FOFF1}       & \multicolumn{1}{c|}{-0.214$^{***}$}     & \multicolumn{1}{c|}{-0.190$^{***}$}                  & \multicolumn{1}{c|}{0.416$^{***}$}       & \multicolumn{1}{c|}{-1.205$^{***}$} & -1.233$^{***}$ \\
\multicolumn{1}{c|}{}            & \multicolumn{1}{c|}{(0.031)}            & \multicolumn{1}{c|}{(0.035)}                         & \multicolumn{1}{c|}{(0.087)}             & \multicolumn{1}{c|}{(0.182)}        & (0.200)        \\ \hline
\multicolumn{1}{c|}{LEISURE}     & \multicolumn{1}{c|}{0.097$^{***}$}      & \multicolumn{1}{c|}{0.118$^{***}$}                   & \multicolumn{1}{c|}{0.026}               & \multicolumn{1}{c|}{0.544$^{***}$}  & 0.638$^{***}$  \\
\multicolumn{1}{c|}{}            & \multicolumn{1}{c|}{(0.030)}            & \multicolumn{1}{c|}{(0.034)}                         & \multicolumn{1}{c|}{(0.092)}             & \multicolumn{1}{c|}{(0.172)}        & (0.192)        \\ \hline
\multicolumn{6}{c}{Significance at the 1\% (***), 5\% (**), and 10\% (*) levels.}                                                                                                                                                  
\end{tabular}
\caption{ZTNB and OIZTNB models fit to the fishing trip data set.}
\label{tab:Alaska}
\end{table}
}

In ``Models for truncated counts'' \citep{grogger1991models}, zero-truncated Poisson and negative binomial (PP and ZTNB) models are presented as important and viable in the fields of recreation demand, crime, transporation, and labour economics. The merits of the truncated models (in comparison to the untruncated counterparts), in particular the merits of ZTNB, are exemplified. To illustrate, the authors estimate a participation equation using data on the number of fishing trips taken by households in Alaska.

Estimating an OIZTNB model, average absolute one-inflation is estimated to be 4.4\%, and average inflation is -2.2\%. The Wald statistic (p-value 0.000) supports the presence of one-inflation, but the LRT is borderline (p-value is 0.09, unless the insignificant variables are dropped from $Z$, in which case the p-value is 0.02). The fit of the models is illustrated in Figure \ref{fig:Alaska}, and estimated parameters are reported in Table \ref{tab:Alaska}. Most notably, a significant marginal effect (on \texttt{AVLONG}) changes sign.

Our replication efforts use data from \cite{carson2009nested}, with small differences in the parameter estimates from ZTNB, due to our inability to reproduce a \texttt{TRATE} variable that was an inclusive variable from a nested logit model. Nonetheless, the highlighted discrepancies between the ZTNB and OIZTNB models should be invariant to the presence of this missing variable; when one-inflation is present (as supported by the Wald and LRT), additional variables can not explain away its presence unless they are linked to the one-inflation parameter.

\section{Conclusions}

In administrative settings, the number of visits is typically a truncated count. For such data, the individual is not observed unless they visit the entity that is collecting the data. For example, data on the number of health care visits typically begin at the 1-count; individuals that do not visit the doctor are not observed. A common endeavour is to determine how a policy or characteristic might influence the expected number of visits, and for data that are zero-truncated, an appropriate zero-truncated distribution must be used in order to avoid biased and inconsistent estimators.

The current standard truncated count model, the zero truncated negative binomial (ZTNB), is likely biased and inconsistent in most settings that it is employed. This is due to one-inflation. One-inflation occurs because the individual experiences being observed during their first visit. Information is revealed, and behaviour is altered, typically in a way that discourages future visits. Individuals must experience the observational setting before they can enter the data set, but the experience itself alters their behaviour. This causes altered 1 counts as the experience either encourages or dissuades future visits. While one-inflated distributions have been developed in areas outside social sciences, the distributions have not yet been extended to a regression framework, until now.

This paper develops one-inflated zero truncated negative binomial (OIZTNB) and one-inflated positive Poisson (OIPP) regression models, as well as a plethora of related estimators and tools, all made available in the \texttt{R} package \texttt{oneinfl}. The methods developed here are applied to four data sets that have traditionally been used to champion the ZTNB regression model. In all cases, one-inflation is found, suggesting that the ZTNB model is actually misspecified. Simulation studies suggest large biases when specifying the wrong model. Accordingly, we recommend the use of the OIZTNB model in place of the ZTNB, for administrative type data.

\bibliographystyle{apa}
\bibliography{refs}

\newpage

\begin{appendix}

\section{Log-likelihoods} \label{app:logl}

\subsection{OIPP model}

From equation \ref{eq:OIPP}, the log-likelihood for the one-inflated positive Poisson (OIPP) model is:
\begin{equation}
\begin{split}
\ell = &\sum_{i = 1}^n \Bigg\{ \log (1 - \omega_i) + I_1 \log \left[\frac{\omega_i}{1-\omega_i}+\frac{\lambda_i}{\exp{\lambda_i}-1}\right]\\ 
&+ \left(1-I_1\right)\left[y_i \log{\lambda_i} - \log{\left[\exp{\lambda_i}-1\right]} - \log{y_i!}\right] \Bigg\}
\end{split}
\label{eq:loglOIPP}
\end{equation}

\noindent where $I_1 = 1$ if $y_i = 1$ and 0 otherwise, and where $\lambda_i$ and $\omega_i$ have been previously defined.

\subsection{OIZTNB model}

From equation \ref{eq:OIZTNB}, a log-likelihood function for the OIZTNB model is:
\begin{equation}
\begin{split}
\ell = &\sum_{i = 1}^n \Bigg\{ \log (1 - \omega_i) + I_1 \log \bigg[\Big(\frac{\omega_i}{1 - \omega_i} + \alpha\Big(\frac{\alpha}{\alpha + \lambda_i}\Big)^\alpha\lambda_i\Big)/\Big(\alpha + \lambda_i - \alpha\Big(1 + \frac{\lambda_i}{\alpha}\Big)^{1-\alpha}\Big)\bigg]\\
& + (1 - I_1)\bigg[\sum_{j=1}^{y_i} \log (\alpha + j - 1) - \log y_i! + \alpha \log \alpha + y_i \log \lambda_i - (\alpha + y_i) \log (\alpha + \lambda_i)\\
&- \log\bigg(1-\bigg(1+\frac{\lambda_i}{\alpha}\bigg)^{-\alpha}\bigg)\bigg]\Bigg\}
\end{split}
\label{eq:logl}
\end{equation}

\noindent where $I_1$, $\lambda_i$, and $\omega_i$ are defined previously. The recursive property of the gamma function has been invoked in order to eliminate it: \(\log \Gamma(\alpha + k) - \log \Gamma(\alpha) = \sum_{j=1}^k\log(\alpha + j - 1)\).

In both OIPP and OIZTNB models, $\log y_i!$ appears. These terms could be removed for faster optimization, but for the calculation of likelihood ratio test (LRT) statistics. R is unable to calculate $\log y_i!$ for $y_i > 170$, so in \texttt{oneinfl} we use Stirling's approximation ($\log y_i! \approx y_i \log y_i - y_i$) for values that are large.

\section{Marginal effects} \label{app:marginal}

\subsection{Marginal effects for the OIPP model}

The mean of the OIPP distribution is:
\begin{equation}
\mathbb{E}[y_i^{OIPP}] = \omega_i + (1 - \omega_i)\frac{\lambda_i \exp(\lambda_i)}{\exp(\lambda_i) - 1}
\label{eq:OIPPmean}
\end{equation}  

\noindent The first derivative of equation \ref{eq:OIPPmean} with respect to a regressor \(q_{ij}\) in \(X\) and/or \(Z\) is:
\begin{equation}
\frac{\partial \mathbb{E}[y_i^{OIPP}]}{\partial q_{ij}} = \frac{\partial \omega_i}{\partial q_{ij}}\left[1 - \frac{\lambda_i \exp(\lambda_i)}{\exp (\lambda_i) - 1} \right] + \frac{\partial\lambda_i}{\partial q_{ij}} \left(1 - \omega_i\right) \frac{\exp(\lambda_i)\left[ \exp(\lambda_i) - \lambda_i - 1 \right]}{\left\lbrace \exp(\lambda_i) - 1\right\rbrace ^ 2},
\label{eq:marginalOIPP}
\end{equation}

\noindent where
\begin{equation}
\begin{split}
\frac{\partial \omega_i}{\partial q_{ij}} &= \frac{\partial \lambda_i}{\partial q_{ij}} \left\lbrace \frac{\exp(\lambda_i) - \lambda_i \exp(\lambda_i) - 1}{\left[ \exp(\lambda_i) - \lambda_i - 1 \right] ^ 2} \right\rbrace \left[ \frac{- \exp(-\boldmath{Z}_i \gamma)}{1 + \exp(-\boldmath{Z}_i \gamma)} \right] \\ 
& \quad - \frac{\partial\exp(-\boldmath{Z}_i\boldmath{\gamma)}}{\partial q_{ij}} \left\lbrace \frac{\exp(\lambda_i) - 1}{\left[\exp(\lambda_i) - \lambda_i - 1\right]\left[1 + \exp(-\boldmath{Z}_i \gamma) \right] ^ 2} \right\rbrace,\\
\lambda_i &= \exp\left(\boldmath{X}_i\boldmath{\beta}\right),\\
\frac{\partial\lambda_i}{\partial q_{ij}} &=
    \begin{cases}
      \lambda_i\beta_j, & \text{if } q_{ij} \text{ is the } j^{th} \text{ column in } \ \boldmath{X}_i \\
      0, & \text{otherwise}
    \end{cases}, \\
\frac{\partial\exp(-\boldmath{Z}_i\boldmath{\gamma})}{\partial q_{ij}} &=
    \begin{cases}
      -\exp(-\boldmath{Z}_i\boldmath{\gamma})\gamma_j, & \text{if } q_{ij} \text{ is the } j^{th} \text{ column in } \ \boldmath{Z}_i \\
      0, & \text{otherwise}
    \end{cases}.
\end{split}
\nonumber
\end{equation}

\noindent As in most non-linear models, the marginal effects themselves are highly non-linear, and are functions of the values of the regressors themselves. That is, the marginal effects vary by individual. When the impact of a regressor is interpreted, a decision needs to be made regarding values for $\boldmath{X}_i$ and $\boldmath{Z}_i$ in the evaluation of $\partial y_i / \partial q_{ij}$.

\subsection{Marginal effects for the OIZTNB model}

The mean of the OIZTNB distribution is:

\begin{equation}
\mathbb{E}[y^{OIZTNB}_i] = \omega_i + (1 - \omega_i) \left[\frac{\lambda_i}{1 - \left(1 + \frac{\lambda_i}{\alpha}\right)^{-\alpha}}\right]
\label{eq:meanOIZTNB}
\end{equation}

\noindent The first partial derivative of equation \ref{eq:meanOIZTNB} with respect to a regressor \(q_{ij}\) in \(X\) and/or \(Z\) is:

\begin{equation}
\begin{split}
\frac{\partial \mathbb{E}[y^{OIZTNB}_i]}{\partial q_{ij}} &= \frac{\partial \omega_i}{\partial q_{ij}}\left[1 - \frac{\lambda_i}{1 - \left(1 + \frac{\lambda_i}{\alpha}\right)^{-\alpha}}\right]\\
&+ \frac{\partial\lambda_i}{\partial q_{ij}} \frac{(1 - \omega_i)}{1 - \left(1 + \frac{\lambda_i}{\alpha}\right)^{-\alpha}}\left[1 - \frac{\lambda_i \left(1 + \frac{\lambda_i}{\alpha}\right)^{-\alpha - 1}}{1 - \left(1 + \frac{\lambda_i}{\alpha}\right)^{-\alpha}}\right],
\end{split}
\label{eq:marginalOIZTNB}
\end{equation}

\noindent where

\begin{equation}
\begin{split}
\frac{\partial \omega_i}{\partial q_{ij}} &= \frac{\partial L}{\partial q_{ij}} \left[1 - \frac{1}{1 + \exp(-\boldmath{Z}_i\boldmath{\gamma})} \right] - \frac{\partial \exp(-\boldmath{Z}_i\boldmath{\gamma})}{\partial q_{ij}} \left\lbrace \frac{(1 - L)}{\left[1 + \exp(-\boldmath{Z}_i\boldmath{\gamma})\right] ^ 2} \right\rbrace,\\
\frac{\partial L}{\partial q_{ij}} &= -\frac{\partial f_1}{\partial q_{ij}} \left[ \frac{1}{(1 - f_1) ^ 2} \right],\\
\frac{\partial f_1}{\partial q_{ij}} &= \frac{\partial \lambda_i}{\partial q_{ij}} \left(1 + \frac{\lambda_i}{\alpha} \right)^{-\alpha} \left[1 + \frac{\lambda_i}{\alpha} - \left(1 + \frac{\lambda_i}{\alpha}\right)^{1-\alpha}\right]^{-1} \\
& \quad \left\lbrace {1 - \frac{\alpha \lambda_i}{\alpha + \lambda_i} - \frac{1}{\alpha} \left[1 + \frac{\lambda_i}{\alpha} - \left(1 + \frac{\lambda_i}{\alpha} \right) ^ {1 - \alpha} \right] ^ {-1} \left[1 - (1-\alpha)\left(1 + \frac{\lambda_i}{\alpha} \right)^{-\alpha} \right]} \right\rbrace,
\end{split}
\nonumber
\end{equation}

\noindent and where $\lambda_i$ and $\frac{\partial \lambda_i}{\partial q_{ij}}$ are as defined previously, and where $f_1$ is the probability of a 1 count according to equation \ref{eq:OIZTNB}.

\end{appendix}

\end{document}